\documentclass[10pt,conference]{IEEEtran}
\IEEEoverridecommandlockouts
\usepackage{cite}
\usepackage{amsmath,amssymb,amsfonts}
\usepackage{algorithmic}
\usepackage{graphicx}
\usepackage{textcomp}
\usepackage{xcolor}
\def\BibTeX{{\rm B\kern-.05em{\sc i\kern-.025em b}\kern-.08em
    T\kern-.1667em\lower.7ex\hbox{E}\kern-.125emX}}
\begin{document}

\title{Using Quantum Solved Deep Boltzmann Machines to Increase the Data Efficiency of RL Agents}

\author{\IEEEauthorblockN{Daniel Kent}
\IEEEauthorblockA{\textit{Frazer-Nash Consultancy} \\
Manchester, United Kingdom \\
d.kent@fnc.co.uk}
\and
\IEEEauthorblockN{Clement O'Rourke}
\IEEEauthorblockA{\textit{Frazer-Nash Consultancy} \\
Middlesbrough, United Kingdom \\
c.orourke@fnc.co.uk}
\and
\IEEEauthorblockN{Jake Southall}
\IEEEauthorblockA{\textit{Frazer-Nash Consultancy} \\
Middlesbrough, United Kingdom \\
jake.southall@fnc.co.uk}
\and
\IEEEauthorblockN{Kirsty Duncan}
\IEEEauthorblockA{\textit{Frazer-Nash Consultancy} \\
Bristol, United Kingdom \\
kirsty.duncan@fnc.co.uk}
\and
\IEEEauthorblockN{Adrian Bedford}
\IEEEauthorblockA{\textit{OxbrdgRbtx} \\
Stratford-upon-Avon, United Kingdom \\
adrian@oxbrdgrbtx.com}
}
\IEEEoverridecommandlockouts

\IEEEpubid{\makebox[\columnwidth]{\begin{tabular}[l]{@{}l@{}}\copyright 2024 IEEE.  Personal use of this material is permitted.  Permission from\\IEEE must be obtained for all other uses, in any current or future media,\\including reprinting/republishing this material for advertising or promotional\\purposes, creating new collective works, for resale or redistribution to servers\\or lists, or reuse of any copyrighted component of this work in other works.\end{tabular}} \hspace{\columnsep}\makebox[\columnwidth]{ }}

\maketitle

\IEEEpubidadjcol

\begin{abstract}
Deep Learning algorithms, such as those used in Reinforcement Learning, often require large quantities of data to train effectively. In most cases, the availability of data is not a significant issue. However, for some contexts, such as in autonomous cyber defence, we require data efficient methods. Recently, Quantum Machine Learning and Boltzmann Machines have been proposed as solutions to this challenge. In this work we build upon the pre-existing work to extend the use of Deep Boltzmann Machines to the cutting edge algorithm Proximal Policy Optimisation in a Reinforcement Learning cyber defence environment. We show that this approach, when solved using a D-WAVE quantum annealer, can lead to a two-fold increase in data efficiency. We therefore expect it to be used by the machine learning and quantum communities who are hoping to capitalise on data-efficient Reinforcement Learning methods.
\end{abstract}


\section{Introduction}
In recent years, Reinforcement Learning (RL) methods are being increasingly deployed as a viable means of defence against rapidly evolving cyber adversaries~\cite{dhir_2021}. For example, with RL it is possible to train autonomous defence agents that can defend a cyber network against malicious attacks ~\cite{ridley_2018, andrew_2022}. However, a challenge in this domain is that it is occasionally not possible to fully train agents due to the restricted access of training data. This can occur due to many reasons, such as operational importance, safety constraints, or insufficient adversary data. In these scenarios, we require data efficient learning methods, capable of fully training agents that have limited access to training data. Therefore, there is a clear need for RL agents that can learn with limited data.

Early research indicates that Quantum Machine Learning (QML) offers a potential route to train ML models with less data than using classical (\textit{i.e.} non-quantum) computing~\cite{huang_2021, dong_2008, pozza_2022}. A key reason for this is because QML can allow data efficient models to be fully trained in timescales that would otherwise be infeasible using classical methods. For example, researchers have shown that maze traversal problems can be solved more efficiently using quantum trained Deep Boltzmann Machines (DBMs), compared to classically equivalent Restricted Boltzmann Machines (RBM)~\cite{crawford_2018}. Here quantum mechanics was used in the training of classical Boltzmann models, giving them the name quantum-hybrid models. However, although a step in the right direction, the current research is limited in a few regards. First, it is often based on quantum simulations rather than on a realised quantum device, meaning that it is unclear if we achieve a quantum advantage in situations that are currently infeasible to simulate classically. Second, RL algorithms used to test quantum advantage are often early RL methods, such as Q-Learning \cite{pozza_2022}, which are regularly outperformed by cutting-edge RL algorithms such as Proximal Policy Optimisation (PPO) \cite{schulman_2017}. Therefore, it is unclear if the quantum advantage seen in early RL algorithms carry over to cutting-edge methods that are preferred in industry. Last, the research is often carried out using bespoke code, rather than leading RL libraries such as StableBaselines3 (SB3)\cite{raffin_2021} that are regularly updated to include all the latest performance enhancement methods. This paper addresses all of these limitations by integrating DBMs, trained using the D-Wave quantum annealer, into SB3 PPO agents. 

In practice, this novel integration was carried out by replacing the default neural networks within SB3 PPO agents with quantum-hybrid DBMs. This allowed for the benefits of state-of-the-art algorithms and leading python libraries to be combined with the benefits of a DBM, as well as a direct comparison of quantum trained DBMs to feed-forward trained neural networks. The RL agents were evaluated using PrimAITE, Dstl’s primary cyber defense testing environment for artificial intelligence. This allowed us to test our work in sophisticated scenarios commonly used by the cyber defense community.  With this process, we were able to test the hypothesis:

``Quantum trained DBMs can be more data efficient, without a loss of accuracy, than classically trained neural networks when used within SB3 PPO agents.''

We found that quantum-hybrid DBM approach can lead to a two-fold increase in data efficiency over SB3 PPO agents. Thus showing an emergent result of how QML can increase the data efficiency of cutting-edge cyber defence agents.

\section{Background}\label{section:background}
\subsection{Reinforcement Learning}\label{section:rl}
Reinforcement Learning is a subset of Machine Learning (ML). Its primary aim is to train intelligent agents to take the actions in an environment that maximise a user defined reward. How the agent selects actions is known as the policy \cite{sallans_2004}. In other words, the policy, $\pi (s)$, is a probability distribution over available actions given a state, $s$. From the latest observation of the environment at time $t$, the agent is in the state $s_t$, receives a reward $r_t$, and chooses an action $a_t$ that affects the environment and moves it into a new state. This transition to a new state $s_{t+1}$ generates a new reward $r_{t+1}$ for the agent who then chooses a new action $a_{t+1}$. 

By acquiring rewards over time, the agent learns the best actions to choose given a certain state, with the goal of moving the environment into a new state that leads to the maximum reward. Through maximising the cumulative reward received, the agent can find the optimal policy and therefore maximise its performance. This is the goal of RL and we often employ three classes of algorithms to learn the optimal policy.

The first are known as value based methods, which is where the agent attempts to learn either the \textit{state value} function, $V_{\pi}(s)$, or the \textit{state-action} value function, $Q_{\pi}(s,a)$. The state value is the cumulative reward of being in a state, $s$, and following a particular policy, $\pi$. The state-action value is the cumulative reward of being in a state, $s$, taking an action, $a$, and then following a particular policy thereafter. Using this information, the agent can then update it's policy to attain maximum reward. For example, its policy could be to select the action that gives the greatest state-action value. This is the main principle of Q-Learning.

The second are known as policy based methods where the agent attempts to learn the optimal policy directly. Policy based methods attempt to maximise an agent's reward through an iterative process of policy improvement. This is achieved through evaluating the current policy used by the agent, generating new policies, and then comparing their performance and selecting the one that will maximise the reward. For example, one may utilise policy gradient methods to move a policy closer to the optimal policy \cite{sutton_2018}.

The final class of methods are known as actor-critic methods. This is where the agent utilises both value and policy based methods to learn the optimal policy. An example of a cutting-edge actor-critic method is PPO \cite{schulman_2017}. 

Often, these value and policy functions are parameterised by neural networks. For example, if the state space is large we can approximate the state-action function with a set of parameters, $\Theta$, so that $V_{\pi}(s)\approx V_{\pi}(\Theta)$. These $\Theta$ parameters are then used as inputs to a neural network with the output being equal to the state value. Learning is then achieved through updating the weights of this neural network. For value function networks, the output layer will be a single node and its activation represents the scalar value function. For policy function networks, the output layer will have multiple nodes where each node's probability at activation represents the probability of taking a particular action. 

\begin{figure*}
	\centering
	\includegraphics[width=\linewidth]{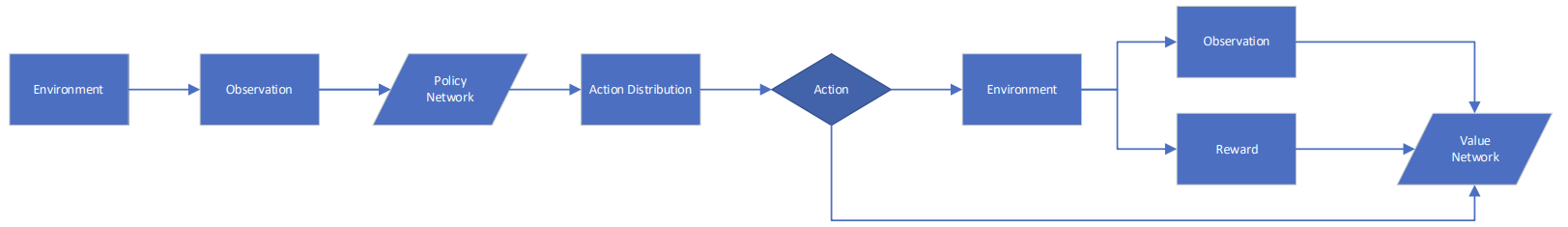}
	\caption{Visualisation of how actions are chosen by the policy network, evaluated by the environment, and then used as inputs to the value network. This gives a value function estimate for a given action, observation and reward. In this work we use DBMs for both the policy and value networks.} 
	\label{fig:ppo_flow}
\end{figure*}

In this work we implement PPO agents using SB3 \cite{raffin_2021}, a leading python package of this cutting-edge algorithm. SB3 offer functionality to replace their default neural networks with customised networks. We utilise this to replace their networks with DBMs. We do not give the full details of the mathematics behind SB3's PPO implementation because this can be found in the references and would diverge from the key message of this paper. However, we do present the flow of how PPO calculates its value network estimate in Fig. \ref{fig:ppo_flow}. The main points we emphasis are:
\begin{enumerate}
	\item We implement PPO using SB3, a leading library of RL algorithms.
	\item SB3's PPO agents by default use two uni-directional neural networks, one for each of the value and policy functions. We will replace each and both of these neural networks with DBMs and report the difference in data efficiency.
\end{enumerate}

We train these DBMs using the D-Wave Advantage system, which is a quantum device that consists of over $5000$ qubits. It utilises various quantum phenomena to solve optimisation problems via quantum annealing \cite{dwave_doc_getting_started}.

\subsection{Boltzmann Machines}\label{section:bm}
\begin{figure}
	\centering
	\includegraphics[width=0.7\linewidth]{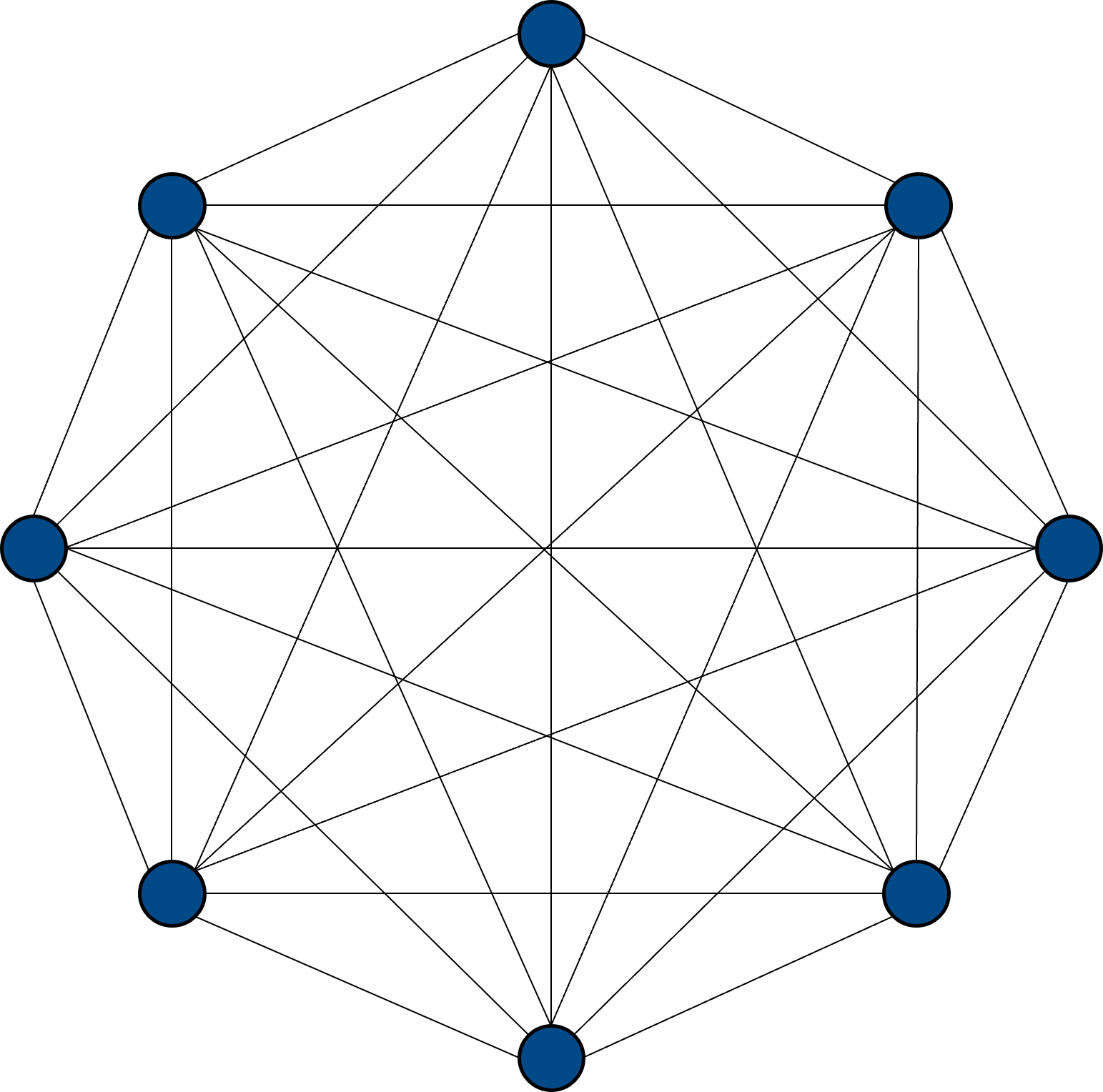}
	\caption{Visualisation of the connections between units in a Boltzmann Machine. The output of this machine is the energy defined by \eqref{eq:bm}.}
	\label{fig:boltzmann_machine_figure}
\end{figure}

Boltzmann Machines, see Fig. \ref{fig:boltzmann_machine_figure}, are undirected networks, defined by a set of units $u_i$ for $i=1,...,N$, weights between the units $w_{ij}$ for $i,j = 1, ... , N$ and biases $a, b_i$
for $i = 1, ... , N$. Generally, the units are binary variables, i.e. they equal $1$ or $0$, and this the approach we take in this paper. For a given state, the energy of the Boltzmann Machine is given by \cite{pochart_2022}
\begin{equation}\label{eq:bm}
	E(\mathbf{u})=a+\sum_{i=1}^{N}b_iu_i+\sum_{i=1}^{N}\sum_{j=i+1}^{N}w_{ij}u_iu_j\,.
\end{equation}
We often refer to this energy function of a Boltzmann Machine as its Hamiltonian. The probability of a Boltzmann machine being in a certain state is given by the Boltzmann distribution. This distribution is defined by
\begin{equation}\label{eq:bmdist}
	P(\mathbf{u})=\frac{1}{Z}e^{-\beta E(\bf{u})}\,,
\end{equation}
where $\beta$ is some fixed thermodynamic constant and $Z=\sum_{\bf{u}}e^{-\beta E(\mathbf{u})}$ is a normalising constant. This normalising constant is known as the partition function, which ensures the probabilities sum to $1$. Therefore, states with the lowest energy have the highest probability of occurring.

\begin{figure}
	\centering
	\includegraphics[width=0.85\linewidth]{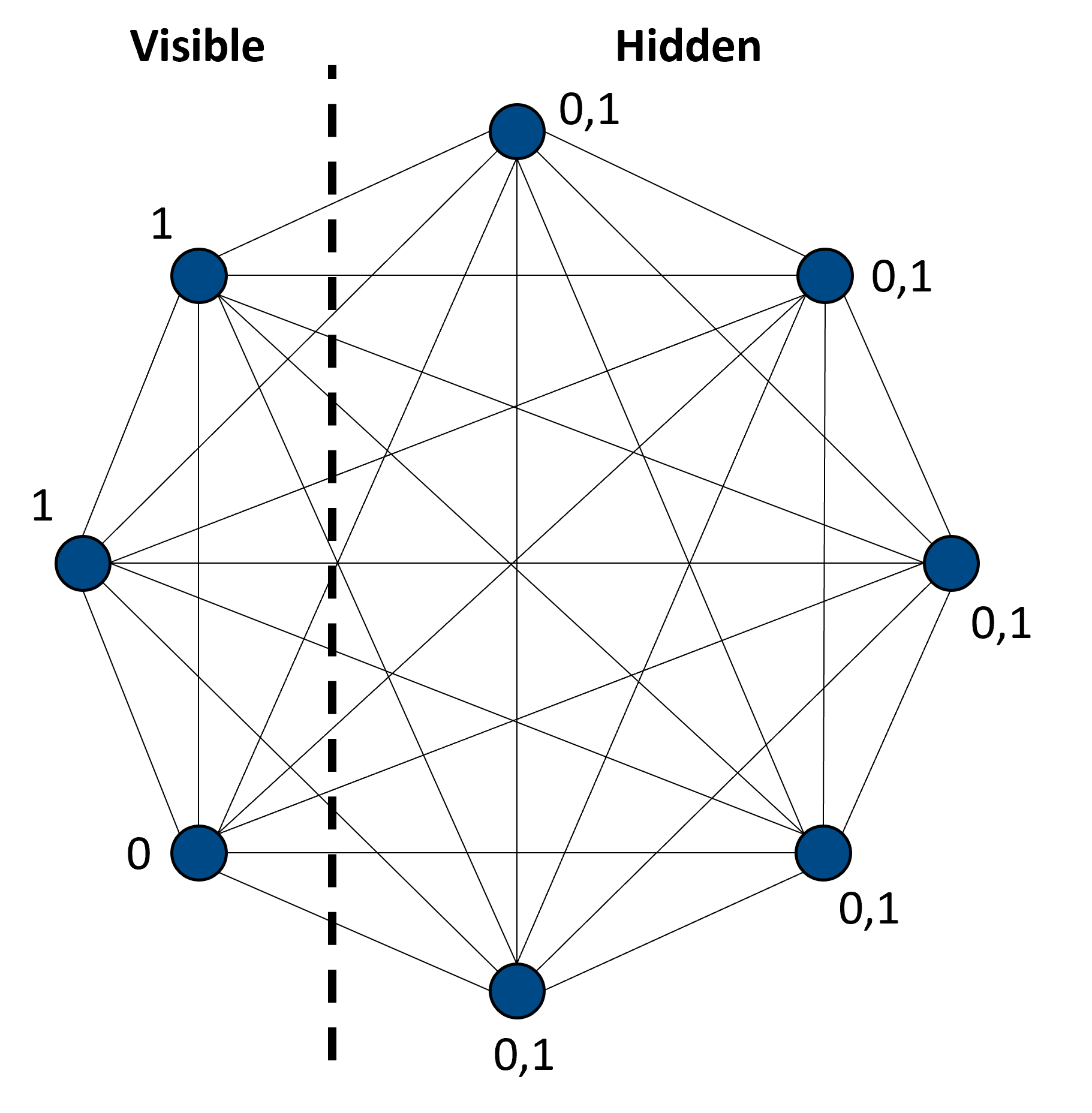}
	\caption{Visualisation of the connections between units in a clamped Boltzmann Machine. The visible units are clamped to either $0$ or $1$, representing a fixed input and output, but the hidden units can vary. In this work, the output of this machine is the energy defined by \eqref{eq:free_energy}.}
	\label{fig:clamped_boltzmann_machine_figure}
\end{figure}

We can split the units of a Boltzmann Machine into visible units, $\mathbf{v}$, and hidden units, $\mathbf{h}$, so that $\mathbf{u}=\left(\mathbf{v}, \mathbf{h}\right)$. Here, the visible units represent the inputs and outputs of a system, whereas the hidden units are only used during computation. In this work, we use clamped Boltzmann Machines. This means we sample the energy of the system with the visible units clamped, representing a fixed input and output sample from the system being approximated, see Fig. \ref{fig:clamped_boltzmann_machine_figure}. Therefore, $E$ and $P$ are approximated via the equilibrium free energy of the fixed state, $\mathbf{v}$, which is equal to the average energy of the clamped system plus an entropy term. This quantity is defined by the equation
\begin{equation}\label{eq:free_energy}
	F(\mathbf{v}):= \sum_{\mathbf{h}}P(\mathbf{v,h})E(\mathbf{v,h})+\frac{1}{\beta}\sum_{\mathbf{h}}P(\mathbf{v, h})\log(P(\mathbf{v}, \mathbf{h}))\, ,
\end{equation}
where $E$, $F$ and $P$ are written as functions of $\mathbf{v}$ and $\mathbf{h}$, and $\beta$ is a fixed thermodynamic constant \cite{sallans_2004}. 

\begin{figure}
	\centering
	\includegraphics[width=\linewidth]{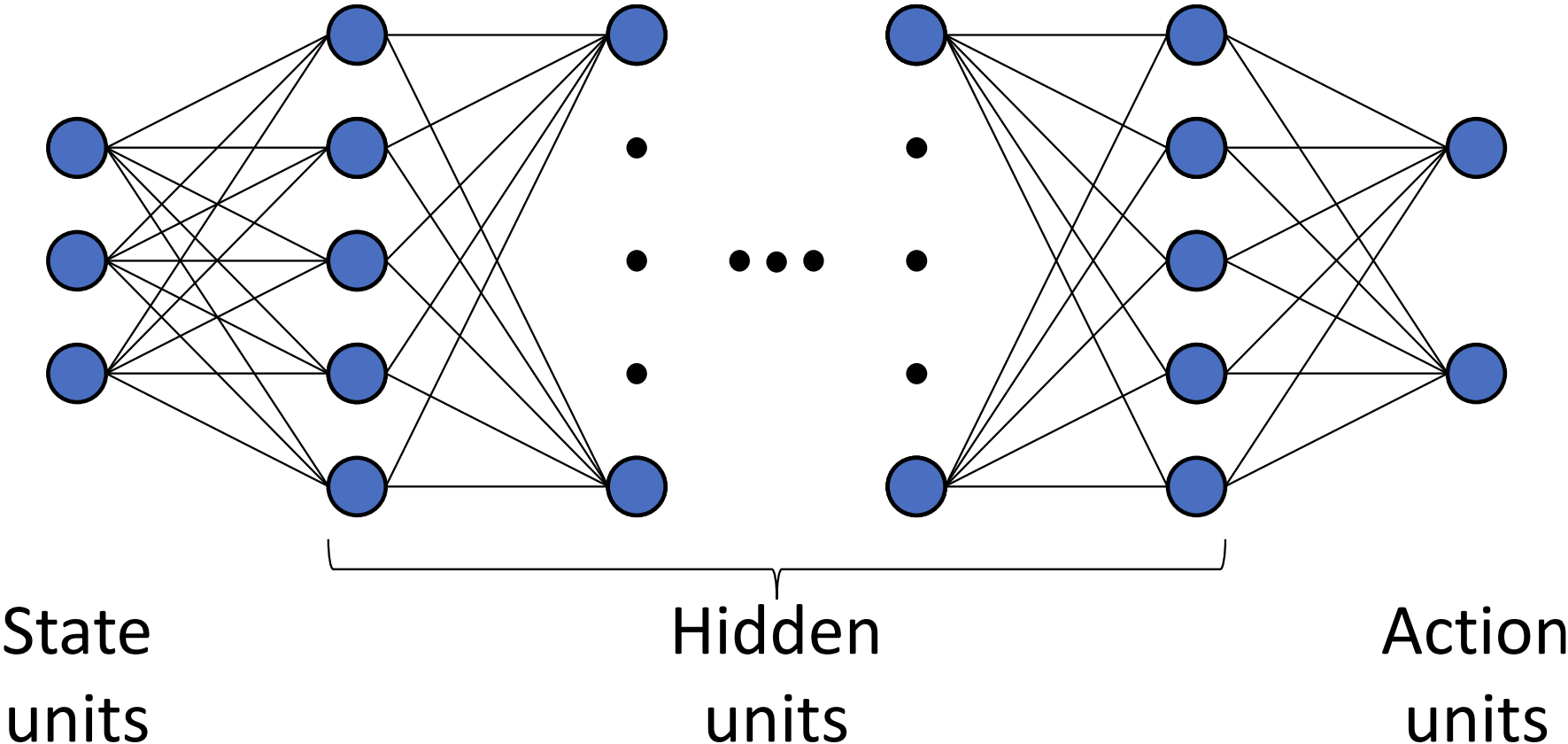}
	\caption{Visualisation of the connections between units in a DBM. In this work, the output of this machine is the energy defined by \eqref{eq:free_energy}.}
	\label{fig:dbm}
\end{figure}

A DBM further splits the visible and hidden units into subgroups: the visible units are split into `state units’ and ‘action units’ ($\mathbf{s}$ and $\mathbf{a}$, respectively, so that $\mathbf{v}=\left(\mathbf{s}, \mathbf{a}\right)$), and the hidden units are split into distinct layers $\mathbf{h}_1,...,\mathbf{h}_l$. The state units are connected to the $\mathbf{h}_1$ units, the $\mathbf{h}_i$ units are connected to the $\mathbf{h}_{i+1}$ units for $i=1,..., l-1$, and the $\mathbf{h}_l$ units are connected to the action units. We show this structure in Fig. \ref{fig:dbm}, and see that it resembles the shape of a typical neural network. However, the connections between units are not directional. Each unit is therefore dependent on every other unit, which makes explicit computation of the free energy intractable for DBMs with more than one hidden layer and large numbers of hidden units.

Previous work by Crawford et al. \cite{crawford_2018} has shown that DBMs provide an advantage over classically equivalent networks, in that they require less data. In their work they approximated Q functions using the negative free energy of Boltzmann Machines, $Q(\mathbf{s},\mathbf{a})\approx-F(\mathbf{s},\mathbf{a})$, followed by updating the Boltzmann Machine weights using a gradient descent method. Note that set of visible units comprise the state and action units.

\subsection{PrimAITE}\label{section:primaiteback}
This work uses the PrimAITE training environment that was developed by QinetiQ within the autonomous resilient cyber defence (ARCD) programme \cite{primaite_2023}. It is named so because it is ARCD’s Primary-level AI Training Environment, providing an effective simulation capability for the purposes of training and evaluating AI in a cyber-defensive role. It includes features that allow modelling of cyber defense scenarios, such as firewall rules, network traffic, and benign users known as Green agents. PrimAITE comes with prepackaged networks that represent example cyber attacks, such as a distributed-denial-of-service (DDoS) attack. Lastly, it also comes pre-packaged with cutting-edge classical algorithms, such as PPO \cite{schulman_2017}, that are implemented by well documented and widely used RL packages such as SB3 \cite{raffin_2021}. Thus, this environment allows us to benchmark our quantum-hybrid approach against leading classical methods.

Scenarios in PrimAITE consist of a network of nodes, representing network endpoints such as servers, and links, representing connections between nodes. In PrimAITE, Red agents can attack the network by either changing the statuses of nodes or preventing benign Green agent network traffic.

To defend against these Red agent attacks, PrimAITE trains a defensive Blue agent using a given RL algorithm. Scenarios are run in episodes made up of discrete timesteps in which the Blue agent first takes an observation of the statuses of network endpoints and traffic. Based on this, the Blue agent takes an action, which could be to repair a damaged network endpoint or creating a firewall. After an action is taken, a reward is returned to the Blue agent based on the resulting state of the cyber network. The underlying RL algorithm then updates the Blue agents policy and/or value function using this reward. In this work, we use PPO as Blue agents underlying RL algorithm, which PrimAITE natively implements using SB3.


\section{Methodology}\label{section:methodology}

Explicit calculation of a DBM’s free energy is not tractable for larger networks. Therefore, we used sampling methods to approximate the free energy\footnote{To find the minimum or average energies of a Boltzmann distribution via sampling is classically NP-hard \cite{barahona_1982}.}. The D-Wave systems are designed for optimisation of Ising or QUBO problems, i.e. finding the minimum energy solution for a given problem. However, the sample set returned is in fact a sample set from a Boltzmann distribution \cite{bian_2010}. We sample the distribution in four steps:
\begin{enumerate}
	\item Define the clamped Boltzmann Machine Hamiltonian as a function of hidden units only, applying additional bias as required to account for connections with visible nodes.
	\item  Sample the energy states of the Hamiltonian using the D-Wave, returning 100 states\footnote{We found the choice of $100$ states to be sufficient to achieve adequate learning without using too much QPU time.}. These states will generally be the lowest energy states, which are the highest probability states.
	\item Calculate the energy (and therefore probability) of all unique samples explicitly. Assume all other states have zero probability of occurring.
	\item Calculate the average free energy using \eqref{eq:free_energy} and use this value as the output of the Boltzmann Machine.
\end{enumerate}

Within StableBaseline3 PPO agents, neural networks have a scalar output for value functions and a vector output for policy functions (a probability for each available action). Therefore, the steps to replace either with a DBM differ. Additionally, using a DBM to model a function with a vector output appears to be a novel application.

The value network in PPO only considers the current state of the network and outputs a single value. Therefore, there are no action units in the DBM and we define the clamped Boltzmann Machine based on the fixed state units, with the hidden units allowed to vary.

The policy network in PPO considers the current state of the network and outputs a value for each possible action, corresponding to that action’s relative probability. This is achieved by defining a unique clamped Boltzmann Machine for each action, in which the relevant action unit has a value of $1$, and the remaining action units have values of $0$. To sample each unique Boltzmann Machine individually would multiply the required wall-clock and QPU time required by the total number of actions, which is infeasible for this work. To avoid this, a further level of approximation is added to the system, and D-Wave sampling is instead performed as follows:
\begin{enumerate}
	\item Define individual Hamiltonians $H_i$ for each action, $a_i$, as functions of hidden units only, applying additional bias as required to account for connections with visible nodes.
	\item  Define a sampling Hamiltonian as the mean over all individual Hamiltonians, $H_{sample}=\frac{1}{n}\sum_{i=1}^{n}H_i$ where $n$ is the number of possible actions.
	\item Sample the energy states of the sampling Hamiltonian using the D-Wave.
	\item For each action, calculate the energy (and therefore probability) of all unique samples explicitly. Assume all other states have zero probability of occurring.
	\item Calculate each action's average free energy using \eqref{eq:free_energy}.
\end{enumerate}

SB3 PPO agents implement PyTorch to construct their neural networks. Therefore, to replace their neural networks with DBMs, we constructed DBM value and policy network python classes that inherit from PyTorch's torch.nn module and make use of the library's functionality. Following this, we then used SB3's custom value and policy network features to replace the default neural networks with DBMs. Weights within the DBM were therefore tracked within the PyTorch python package. This allows automatic calculation of gradients, subsequent weight updates and adaptive learning rates to be handled by existing python libraries.

\begin{figure}
	\centering
	\includegraphics[width=\linewidth]{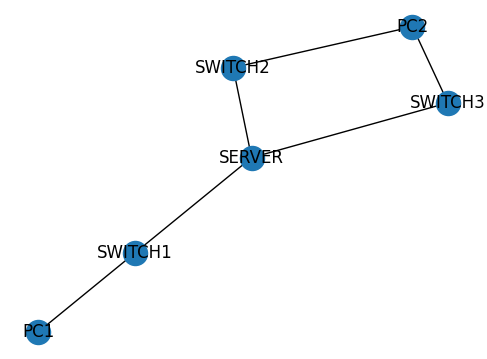}
	\caption{PrimAITE's pre-packaged six-node network that we used in this work. Red agent attacks originated in either of the PC network endpoints and had the ultimate goal of sabotaging the server endpoint.}
	\label{fig:six_node}
\end{figure}

We trained our agents on was PrimAITE's pre-packaged six-node cyber network that represented a DDoS attack, see Fig. \ref{fig:six_node}. This cyber network contained hard-coded Red agents, but we also introduced an additional Red agent to increase the number of episodes it took an agent to learn. This environment had an action space of 2500, and an observation space of 850. This large action space results from the Blue agent being able construct 30 firewalls.
The DBMs we used had two hidden layers of $64$. This is the same size as the neural networks we used, which allowed us to compare equivalently sized uni-directional neural networks and DBMs. 

\section{Results}\label{section:results} 
The key hyperparameter we found that affected performance was PPO's learning rate. We performed learning rate searches for both DBMs and neural networks. In both cases, $3\rm{E}$-$4$ appeared to be the most effective value.

\begin{figure}
	\centering
	\includegraphics[width=\linewidth]{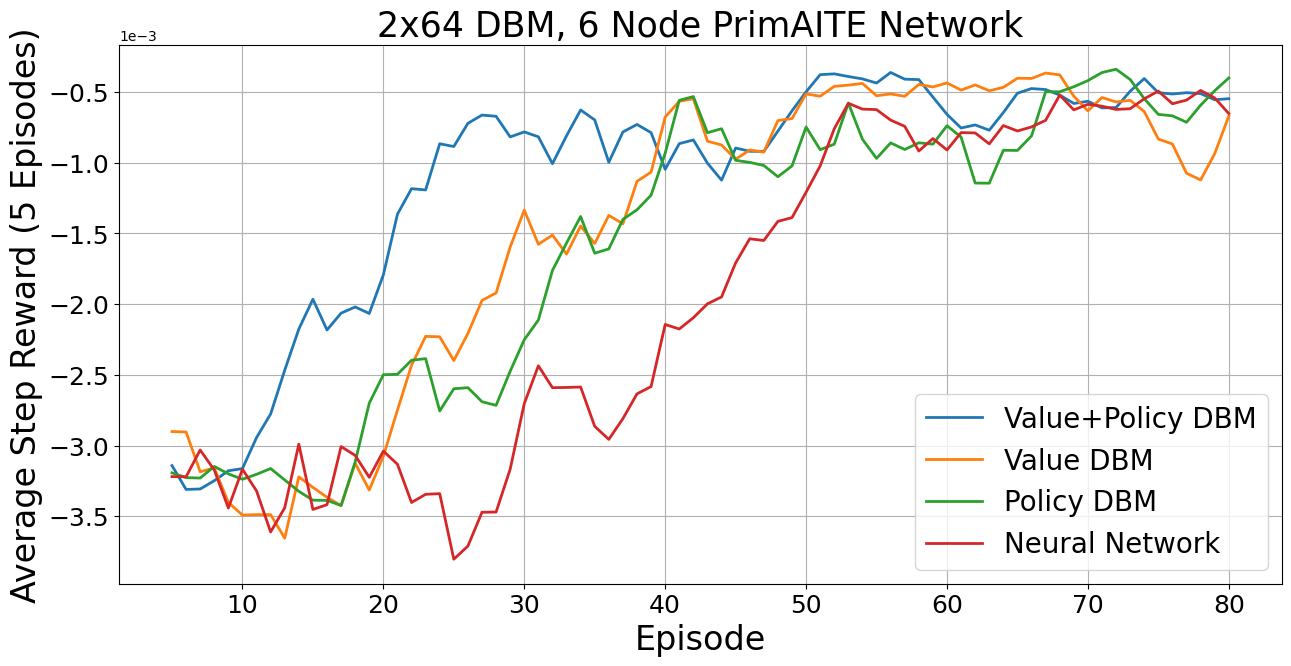}
	\caption{Comparison of DBM and neural network results. For clarity, we averaged rewards over five episodes.}
	\label{fig:results_2xspeedup}
\end{figure}

Figure \ref{fig:results_2xspeedup} shows results from four cases; one with neural networks for both networks, one with a DBM policy network and neural network value network, one with a DBM value network and neural network policy network, and one with a DBM for both the value and policy network. We ran all setups with the same random seed, with additional seed exploration limited by time constraints. However, due to the inherent stochastic nature of sampling the D-WAVE, you cannot guarantee the same results every time.  

\begin{figure}
	\centering
	\includegraphics[width=\linewidth]{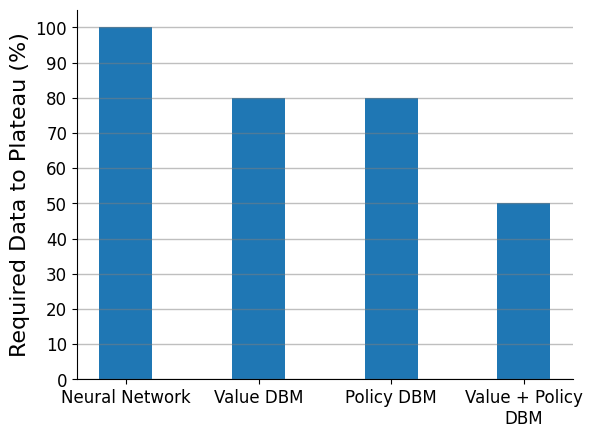}
	\caption{Bar plot showing roughly how many episodes each model takes to plataou, as a percentage of the number of episodes it takes the wholly neural network model to learn.}
	\label{fig:bar_plot}
\end{figure}

The results show all networks reaching a similar plateau, with the neural network case reaching the plateau after approximately $55$ episodes, the single DBM cases reaching the plateau after approximately $40$ episodes, and the dual DBM case reaching the plateau after approximately $25$ episodes. Therefore, the inclusion of a DBM for either network meant that only $80$\% of the data required by a classical agent was needed. For the dual network, this percentage fell to $50$\%, effectively doubling the data efficiency. We show this in Fig. \ref{fig:bar_plot}.

The key takeaway from this section is that when compared on a like-for-like basis, DBMs can be more data efficient that neural networks. However, on average the DBMs took around one hour to train one episode. In contrast, it took the neural networks around five seconds using 16GB RAM. These results therefore suggest a trade-off between data efficiency and wall-clock time to train. However, with improvements in D-WAVE technology, this gap in time and cost resources may decrease over the coming years.

\section{Conclusions}\label{section:conclusion}
This work shows that a quantum hybrid DBM method can lead to a two-fold increase in data efficiency for cyber defence problems, compared to classically trained neural networks. We achieved this by using the novel application of integrating DBMs within SB3's PPO agents, with training carried out using the D-WAVE quantum annealer. However, due to the high cost and time resources, we suggest that the hybrid approach is only suitable where data efficiency is highly prized. Future work includes quantifying the statistical data efficiency of DBMs and neural networks. This could be carried out by training the same network with different random seeds, and then analysing the results. Additionally, this work could be extended to transfer learning, multi agent RL systems, and testing on more complex cyber networks.

\end{document}